\documentclass[11pt]{article}
\usepackage{graphicx}
\begin{document}

\title{Enhancing network transmission capacity by efficiently allocating node capability}

\author{Guo-Qing Zhang, Shi Zhou, Di Wang, Gang Yan, and Guo-Qiang Zhang
\thanks{This work is supported by the National Natural
Science Foundation of China under grant no.\,60673168, the Hi-Tech Research and Development Program
of China under grant no.\,2006AA01Z207 and the China Next Generating Internet Project under grant
no.\,CNGI-04-7-1D. S.~Zhou is supported by The Royal Academy of Engineering and EPSRC (UK) under
grant no.\,10216/70.}
\thanks{G.-Qing\,Zhang, D.\,Wang and G.-Qiang\,Zhang are with Institute of Computing Technology, Chinese Academy of Sciences (CAS), and Graduate University of CAS, Beijing, 100190, P.~R.~China
(e-mail:gqzhang@ict.ac.cn, guoqiang@ict.ac.cn). S.\,Zhou is with
Department of Computer Science, University College London, Malet
Place, London, WC1E 6BT, United Kingdom (e-mail:
s.zhou@cs.ucl.ac.uk). G.~Yan is with University of Science and
Technology of China, Hefei, 230026, P.~R.~China. } }

%\markboth{IEEE Communications Letters}{IEEE Communications
%Letters}

\maketitle

\begin{abstract}

A network's transmission capacity is the maximal rate of traffic
inflow that the network can handle without causing congestion. Here
we study how to enhance this quantity by redistributing the
capability of individual nodes while preserving the total sum of
node capability. We propose a practical and effective
node-capability allocation scheme which allocates a node's
capability based on the local knowledge of the node's connectivity.
We show the scheme enhances the transmission capacity by two orders
of magnitude for networks with heterogenous structures.

\end{abstract}

\section{Introduction}

There are different ways to enhance a network's transmission
capacity. A number of routing strategies have been introduced to
route traffic based on the topological properties of the
network~\cite{Yan06,Fortz00} or the real-time load distribution on
links~\cite{Scholz08}. Recently we reported~\cite{Zhang07} that a
network's transmission capacity can be increased by removing a few
links with certain topological properties. These work assume that
all nodes or links are assigned with uniform resources. However this
is not the case in many networks and it is often difficult to change
a network's routing protocol or topology.

One can also enhance a network's transmission capacity by simply
increasing the resources of all nodes or links. It is, however, very
expensive. A more realistic and economical approach is to
redistribute the resources in the network, such that those that
handle higher volumes of traffic load have more resources whereas
those that handle less load have fewer resources. This should be
done under the condition that the total amount of resources
allocated in the network is fixed. This is relevant to the design
and engineering of communication networks where a key goal is to
deploy limited resources in a way to achieve the best network
performance.

Node capability and link bandwidth are two major resources to be
allocated. Whether to redistribute the node capability or the link
bandwidth depends on what is the major cause for congestion in a
network. For example when a new generation of routers are deployed,
the processing power of routers are greatly improved, then the
congestion is mainly caused by the lack of link bandwidth; whereas
when optical fibre replaced cable, link bandwidth increased by a
number of magnitudes, then the congestion is mainly caused by the
lack of node processing power. Over the time these two situations
may happen alternately in communication networks. In this paper we
focus on the allocation of node capability for those networks where
links have sufficient bandwidth. In our future work we will study
the redistribution of link bandwidth.

This paper is organised as follows. In Section II we introduce three
typical communication network topology models: a random
graph~\cite{erdos59}, a scale-free graph~\cite{Barabasi99} and an
Internet-like graph~\cite{Zhou04d}. We also present a simple yet
widely used traffic-flow model based on the shortest-path
routing~\cite{Zhao05}. In Section III we study a number of
degree-based node-capability allocation schemes using simulations
based on the traffic-flow and topology models. We introduce a scheme
which enhances a network's transmission capacity by two orders of
magnitude by allocating a node's capability as a power function of
the node's connectivity. In Section IV we discuss an alternative way
to estimate the optimal power exponent used in our scheme, and
compares our scheme with a previous scheme~\cite{Zhao05} which is
based on the topological property of betweenness~\cite{goh01}.

\section{Background}\label{section:background}

We consider three network models as examples of typical topologies
of computer and communication networks, which are the
Erd\"os-R\'enyi (ER) model~\cite{erdos59}, the Barab\'asi-Albert
(BA) model~\cite{Barabasi99} and the positive-feedback preference
(PFP) model~\cite{Zhou04d}. In graph theory, degree $k$ is defined
as the number of links a node has. The ER model generates random
networks with a Poisson degree distribution, where most nodes have a
degree close to the average degree. The ER model has been used to
describe the structure of LAN and wireless ad hoc networks. The BA
model generates the so-called `scale-free' networks with a power-law
degree distribution, where a few nodes have very large degrees and
the majority nodes have only a few links. Many communication
networks are found to be scale-free~\cite{Faloutsos99} and the BA
model has been used to study the error and attack tolerance of such
networks~\cite{Albert2000}. The PFP model generates a network
structure which is very similar to the Internet at the autonomous
systems level~\cite{Zhou04d,zhou07a}. For each model we generate ten
networks using random seeds to the same numbers of nodes and links.
Table~\ref{table} shows properties of the three models.

\begin{table}
\caption{Network properties and simulation results} \label{table} \centering
\begin{tabular}{l c c c }
\hline\hline  Topology property &  ER &  BA &  PFP \\
\hline
Number of nodes & 4,000 & 4,000 &  4,000 \\
Number of links & 12,000 & 12,000 & 12,000  \\
Degree distribution & Poisson & $\propto k^{-3}$&  $ \propto k^{-2.2}$  \\
Maximum degree & 18 & 156 & 979  \\
Average shortest distance & 4.82 & 4.17 & 3.12  \\
Average clustering coef. & 0.001 & 0.007 & 0.253\\\hline
Node-capability scheme  &  \multicolumn{3}{c}{Critical package-generating rate $\lambda_c$} \\
\hline
$C\propto 1$            & $~~885_{\pm 21}~$ & $ ~~~~57_{\pm 17}~$ &  $ ~~~48_{\pm 9}~~$ \\
$C\propto k $           & $2,616_{\pm 82}~$ & $ 1,289_{\pm 65}~$ &  $ 4,419_{\pm 108}$ \\
$C\propto k^{1.5}$      & $4,319_{\pm 117}$ & $ 2,954_{\pm 302}$ &  $ 1,636_{\pm 83}~~$ \\
$C\propto k^{\alpha^*}$ & $4,319_{\pm 117}$ & $ 3,284_{\pm 241}$ &  $ 5,126_{\pm 177}$ \\
$C\propto B$            & $6,576_{\pm 215}$ & $ 7,604_{\pm 184}$ &  $ 11,592_{\pm 204}$ \\
%
% **** standard deviation ?
%
%$C\propto 1$            & $~~885_{\pm ?}~$ & $ ~~~~57_{\pm 6.2}~$ &  $ ~~~48_{\pm ?}~~$ \\
%$C\propto k $           & $2,616_{\pm ?}~$ & $ 1,289_{\pm ?}~$ &  $ 4,419_{\pm ?}$ \\
%$C\propto k^{1.5}$      & $4,319_{\pm ?}$ & $ 2,954_{\pm ?}$ &  $ 1,636_{\pm ?}~~$ \\
%$C\propto k^{\alpha^*}$ & $4,319_{\pm ?}$ & $ 3,284_{\pm ?}$ &  $ 5,126_{\pm ?}$ \\
%$C\propto B$            & $6,576_{\pm ?}$ & $ 7,604_{\pm ?}$ &  $ 11,592_{\pm ?}$ \\
\hline Optimal exp. $\alpha^*$ in Fig.\,1  & $1.50_{\pm0.05}$ &  $1.40_{\pm0.03}$ &  $1.10_{\pm0.03}$\\
Fitting exp. $\alpha'$ in Fig.\,2 & $1.49_{\pm0.03}$ &  $1.37_{\pm0.03}$ &  $1.11_{\pm0.02}$ \\
\hline\hline
\end{tabular}
\end{table}

In this study we adopt a similar traffic-flow model used
in~\cite{Yan06, Zhang07,Zhao05, Danila06}. For a network with $N$
nodes, at each time step, $\lambda$ packets are generated at
randomly selected nodes. The destination is chosen randomly. A
packet is routed following the shortest path between source and
destination. The shortest-path routing strategy is widely used in
communication networks, such as the Open Shortest Path First (OSPF)
routing protocol. A node $i$ is assigned a capability, $C_i$, which
is the maximal number of packets the node can handle at a time step.
When the total number of arrived and newly created packets is larger
than $C_i$, the packets are stored in the node's queue and will be
processed in the following time steps on a first-in-first-out (FIFO)
basis. If there are several shortest paths for one packet, one is
chosen randomly. Packets reaching their destination are deleted from
the system. As in~\cite{Yan06, Zhang07,Zhao05, Danila06}, node
buffer size in this traffic-flow model is set as infinite as it is
not relevant to the \emph{occurrence} of congestion.

For small values of the packet-generating rate  $\lambda$, the
number of packets on the network is small so that every packet can
be processed and delivered in time. Typically, after a short
transient time, a steady state for the traffic flow is reached
where, on average, the total numbers of packets created and
delivered are equal, resulting in a free-flow state. For larger
values of $\lambda$, the number of packets created is more likely to
exceed what the network can process in time. In this case traffic
congestion occurs. As $\lambda$ is increased from zero, we expect to
observe two phases: free flow for small $\lambda$ and a congested
phase for large $\lambda$, with a phase transition from the former
to the latter at the critical packet-generating rate $\lambda_c$.

In order to measure $\lambda_c$, we use the order parameter~\cite{Arenas01} $ \eta=\lim_{t\rightarrow
\infty}{{\langle \Delta\Theta \rangle} \over {\lambda\Delta t} }$, where $\Theta(t)$ is the total number of
packets in the network at time $t$, $\Delta\Theta=\Theta(t+\Delta t)-\Theta(t)$,  and $\langle\cdots\rangle$
indicates the average over time windows of $\Delta t$. For $\lambda <\lambda_c$ the network is in the
free-flow state, then $\Delta\Theta\approx0$ and $\eta\approx0$; and for $\lambda>\lambda_c$, $\Delta\Theta$
increases with $\Delta t$ thus $\eta>0$. Therefore in our simulation we can determine $\lambda_c$ as the
transition point where $\eta$ deviates from zero.

\section{Enhancing network transmission capacity}

The critical packet-generating rate $\lambda_c$ is used to measure a
network's overall transmission capacity, which is the maximal amount
of traffic flow that a network can handle without causing
congestion. Increasing network transmission capacity is one of the
major goals for network design and engineering.

In the following we study a number of degree-based node-capability
allocation schemes. For each network model, we obtain the network's
critical packet-generating rate $\lambda_c$ by running the
traffic-flow simulation on the ten instance networks of the model.
Table~\ref{table} shows the average and the bounds of $\lambda_c$
for different node-capability allocation schemes. For comparison
purpose without losing generality, we keep the sum of node
capability in all simulations the same, i.e.~$\sum_i{C_i} =
\sum_i{k_i}=2L$, where $L$ is the number of links.

The first scheme is a simplistic baseline case where we  assign all
nodes with a uniform capability, i.e.~$C$ is equal to the average
degree $\langle k\rangle$ (shown in Table 1 as $C\propto 1$). The
second scheme is to allocate a node's capability proportional to its
degree $k$, i.e.~$C\propto k$. The underlying heuristic is that the
larger degree (i.e.~more incoming traffic from neighbors), the more
traffic a node needs to handle. Table 1 shows there is a substantial
increase of $\lambda_c$ for all the networks when node capability
allocation changes from uniform scheme to degree-proportional
scheme. For the ER network, $\lambda_c$ increases three times while
for the PFP model the increase is two orders of magnitude. This
result highlights the importance of respecting the difference in
node degrees when allocating node capability. This is particularly
true for networks featuring a heterogenous structure where node
degrees vary hugely, such as the BA and PFP models.

In the third scheme we further increase the weight of node degree by
assigning a node's capability proportional to its degree raised to
the power of 1.5, i.e.~$C\propto k^{1.5}$. As shown in
Table~\ref{table}, this scheme produces better results for the ER
and BA networks, but it overdoes for the PFP network. This is
because the PFP network has a few nodes with disproportionably large
degrees (see the maximum degree in Table 1). These nodes, although a
small number, take too large a share of the total node capability.
This leaves the majority of nodes, which are poorly connected, with
very little capability and therefore restrains the network overall
transmission capacity. This suggests there is an optimisation
problem as how to achieve the balance between individual and
collective interest.

\begin{figure}
\centering
\includegraphics[width=7cm]{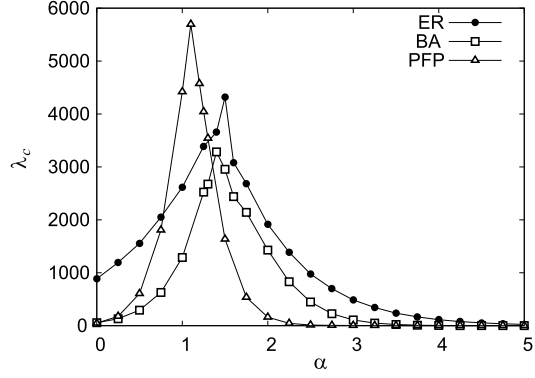}
\caption{Critical
packet-generating rate $\lambda_c$ as a function of exponent
$\alpha$.\label{fig:alpha}}
\end{figure}

In order to examine this issue systematically, we define a generic
degree-based scheme as $C\propto k^{\alpha}.$ When $\alpha=0$ it is
the uniform scheme and when $\alpha=1$ it is the degree-proportional
scheme. Fig.~\ref{fig:alpha} shows the critical package-generating
rate $\lambda_c$ as a function of the exponent $\alpha$. For each
network the $\lambda_c$ peaks at a characteristic value of
$\alpha^*$, which is the optimal exponent for the degree-based
scheme. Table~\ref{table} gives the value of $\alpha^*$ and the peak
value of $\lambda_c$ when using the node-capability allocation
scheme $C\propto k^{\alpha^*}$.

\section{Discussion and Conclusion}
\begin{figure}
\centering
\includegraphics[width=7cm]{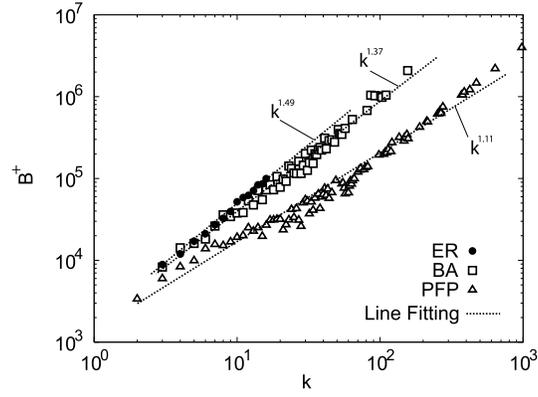}
\caption{Largest betweenness $B^+$ of $k$-degree nodes. The line
fitting is obtained by least square fitting technique.
\label{fig:B-k}}
\end{figure}

For any network topology we can obtain $\alpha^*$ by running the
traffic-flow simulation as described above, which, however, is
time-consuming. Here we introduce an alternative way to estimate
$\alpha^*$ without having to run the simulation. Recently we
analytically proved that a network's critical packet-generating rate
can be estimated as $\lambda_c=min_{\,i\in
V}\{\frac{C(i)N(N-1)}{B(i)}\}$, where $V$ is the set of node
indexes~\cite{Guoqiang07} and $B(i)$ is the betweenness of node $i$
(see definition in ~\cite{goh01}). For the node-capability
allocation scheme of $C\propto k^{\alpha^*}$ we have $\lambda_c
\propto min_{\,k\in K}\{\frac{k^{\alpha^*}N(N-1)}{B^+(k)}\}$ where
$K$ is the set of possible degree values and $B^+(k)$ is the largest
betweenness value of $k$-degree nodes. (We use the largest
betweenness because $\lambda_c$ is constrained by the largest
traffic load a $k$-degree node has.) This suggests that for our
scheme to produce a sound result, $k^{\alpha^*}$ should approximate
$B^+(k)$. As shown in Fig.~\ref{fig:B-k} this is indeed the case
where $B^+(k)\propto k^{\alpha'}$  and $\alpha'\simeq \alpha^*$ (see
Table\,1).  This provides a convenient way to estimate $\alpha^*$ by
fitting the function of $B^+(k)$. If a network has a well-defined
model, the betweenness calculation can be simplified by computing on
a smaller graph generated by the model.

A previous work~\cite{Zhao05} suggested that $\lambda_c$ is
maximised by allocating a node's capability according to the node's
betweenness. As shown in Table~\ref{table} this scheme indeed
produces better results. This is because in the traffic-flow model,
betweenness precisely estimates the traffic load at each node.  This
scheme, however, faces several practical issues. Firstly,
betweenness is sensitive to topology changes. A minor change of a
network's topology, e.g.~adding a new link, could significantly
alter the betweenness value of nodes in all parts of the network.
This requires frequent recalculation of node betweenness for
networks with non-static topologies. Secondly, betweenness
calculation requires global knowledge of a network's topology, which
is not often possible in practice. Finally, betweenness calculation,
the dominant factor of this scheme's processing complexity, is not a
trivial task, especially for large networks. Therefore, the
betweenness scheme is mainly used to give the theoretical upper
bound of $\lambda_c$ for evaluation purpose. While in practice it
can only be used for networks with static topologies, or when the
requirement on network transmission capacity is so high that the
benefit of applying this scheme overruns the cost. By comparison our
degree-based scheme $C\propto k^{\alpha^*}$ is more practical and
robust. Firstly, we allocate a node's capability based on the local
knowledge of node degree. Secondly, the value of $\alpha^*$ is
determined by a network's macroscopic structure which is not
sensitive to minor topology changes, so it can be reused. Finally,
for networks with well-defined model, $\alpha^*$ can be estimated by
doing the fitting on a much smaller network generated by the model,
avoiding betweenness calculation for large real networks, which is
also the dominator factor of this scheme's processing complexity.

In summary we recommend  our node-capability allocation scheme of
$C\propto k^{\alpha^*}$ which can enhance a network's transmission
capacity substantially without increasing the total amount of node
capability. Usually, the value of $\alpha^*$ can be obtained by
fitting the function of $B^+(k)$ without the need to run the
traffic-flow simulation. Our scheme can be viewed as an
approximation of the betweenness scheme with improved practicality
and robustness.

%\bibliography{/10_Work_Bibfile/Ref2008_Oct}

\end{document}